\documentclass[11pt]{article}
\usepackage{amsmath,amsfonts}
\usepackage{graphicx}
\usepackage{epsfig,epsf}
\usepackage{hyperref}

\def\be {\begin{equation}}
\def\ee {\end{equation}}
\def\beq {\begin{equation}}
\def\eeq {\end{equation}}
\def\bea {\begin{eqnarray}}
\def\eea {\end{eqnarray}}

\def\bra {\langle}
\def\ket {\rangle}

\def\beq{\begin{equation}}
\def\eeq{\end{equation}}
\def\barr{\begin{array}}
\def\earr{\end{array}}

\def\gtap{\, \raisebox{-.4ex}{\rlap{$\sim$}} \raisebox{.4ex}{$>$}\,} 

\def\opcit(#1){ {\em op. cit.}, #1}

\def\issue(#1,#2,#3){#1, #2 (#3)} 

\def\APP(#1,#2,#3){Acta Phys.\ Polon.\ \issue(#1,#2,#3)}
\def\ARNPS(#1,#2,#3){Ann.\ Rev.\ Nucl.\ Part.\ Sci.\ \issue(#1,#2,#3)}
\def\CPC(#1,#2,#3){Comp.\ Phys.\ Comm.\ \issue(#1,#2,#3)}
\def\CIP(#1,#2,#3){Comput.\ Phys.\ \issue(#1,#2,#3)}
\def\EPJC(#1,#2,#3){Eur.\ Phys.\ J.\ C\ \issue(#1,#2,#3)}
\def\EPJD(#1,#2,#3){Eur.\ Phys.\ J. Direct\ C\ \issue(#1,#2,#3)}
\def\IEEETNS(#1,#2,#3){IEEE Trans.\ Nucl.\ Sci.\ \issue(#1,#2,#3)}
\def\IJMP(#1,#2,#3){Int.\ J.\ Mod.\ Phys. \issue(#1,#2,#3)}
\def\JHEP(#1,#2,#3){J.\ High Energy Physics \issue(#1,#2,#3)}
\def\JPG(#1,#2,#3){J.\ Phys.\ G \issue(#1,#2,#3)}
\def\MPL(#1,#2,#3){Mod.\ Phys.\ Lett.\ \issue(#1,#2,#3)}
\def\NP(#1,#2,#3){Nucl.\ Phys.\ \issue(#1,#2,#3)}
\def\NIM(#1,#2,#3){Nucl.\ Instrum.\ Meth.\ \issue(#1,#2,#3)}
\def\PL(#1,#2,#3){Phys.\ Lett.\ \issue(#1,#2,#3)}
\def\PRD(#1,#2,#3){Phys.\ Rev.\ D \issue(#1,#2,#3)}
\def\PRL(#1,#2,#3){Phys.\ Rev.\ Lett.\ \issue(#1,#2,#3)}
\def\SJNP(#1,#2,#3){Sov.\ J. Nucl.\ Phys.\ \issue(#1,#2,#3)}
\def\ZPC(#1,#2,#3){Zeit.\ Phys.\ C \issue(#1,#2,#3)}

\begin{document}
\renewcommand*{\thefootnote}{\fnsymbol{footnote}}

\begin{center}
 {\Large\bf{Triplet-extended scalar sector and the naturalness problem}}
 

\vspace{2mm}

Indrani Chakraborty \footnote{indrani300888@gmail.com}
and
Anirban Kundu \footnote{anirban.kundu.cu@gmail.com}

\vspace{3mm}
{\em{Department of Physics, University of Calcutta, \\
92 Acharya Prafulla Chandra Road, Kolkata 700009
}}
\end{center}
\begin{abstract}
We consider the extension of the Standard Model by a complex scalar triplet field,
which occurs naturally in several models of leptogenesis and see-saw mechanism 
for neutrino mass generation, in the context of ameliorating the fine-tuning problem 
of the fundamental scalars through the Veltman Condition, i.e.\ by demanding the 
sum of the quadratically divergent corrections 
to vanish (or be at a reasonable level) by virtue of some possible symmetry of the 
underlying theory. We show that
it is possible to cancel all the scalar one-loop quadratic divergences and hence 
obtain a viable solution for the fine-tuning problem, while satisfying the electroweak 
precision observables, including the $\rho$ parameter, and successfully generating 
the neutrino masses. The stability of the scalar potential puts important constraints on 
the model. 
\end{abstract}

\date{\today}

\noindent PACS no.: {12.60.Fr, 14.80.Ec}

\renewcommand*{\thefootnote}{\arabic{footnote}}
\setcounter{footnote}{0}

\section{Introduction}

Notwithstanding its experimental successes, the Standard Model (SM) is widely believed to be 
an effective theory valid up to a certain scale, above which some new physics (NP) 
takes over. There are several motivations for such an ultraviolet completing NP, e.g., the 
fine-tuning problem of the physical scalar mass, the existence of massive neutrinos and the 
cold dark matter, etc. 

The scalar mass receives a quadratically divergent quantum correction, because 
there is no symmetry to protect the scalar mass, like the gauge symmetry for gauge bosons or 
chiral symmetry for fermions. If the SM is valid up to an energy scale $\Lambda$, the mass of 
the Higgs boson, instead of being $125$ GeV, is expected to be of the order of $\Lambda$, unless
there is an unnatural fine-tuning between the bare mass term and the quantum corrections. The 
standard way out is to appeal to some new symmetry or somehow bring down the Planck scale to 
soften the fine-tuning. 

Let us assume that there is some yet-to-be-discovered symmetry, which protects the scalar mass.
In fact, some well-explored mechanisms, like supersymmetry, may provide this protection, but we 
will be more interested to use a bottom-up approach along with the principle of Occam's razor, 
and try to find the minimal field content that can do the job.
In the framework of cut-off regularization (which, though not Lorentz invariant, is more intuitive 
as this separates the quadratic and logarithmic divergences, and the fine-tuning problem depends 
on the quadratically divergent terms), the sum of all quadratic divergences in the radiative 
corrections to the scalar self energy is set to zero, which is also known as the Veltman condition (VC)
\cite{veltman} \footnote{The sum need not be exactly zero, but should be some small manageable number. 
The higher-order contributions are subleading, so are the contributions of the higher-dimensional 
operators if the Wilson coefficients are perturbative.}. A slightly different VC results if one 
uses dimensional regularization \cite{einhorn}. 
In the SM, the VC can be written in terms of the masses of the Higgs boson, 
the gauge bosons, and the top quark, and, unfortunately, is far from being satisfied; the 
required Higgs mass is more than 300 GeV.

The role of the VC in looking for the possible directions of NP has been well-investigated 
in the literature. 
For example, a possible extension of the SM by one or more gauge singlet scalars satisfying the 
VC, and its possible ramifications in collider searches, as a cold dark matter candidate, or
as a gateway to an ultraviolet complete theory, 
have been discussed in detail in Refs.\ \cite{aksrc,drozd,indrani,masina}. It is easy to check that 
to satisfy the VC for the SM Higgs boson, one needs an extension by bosonic fields that couple to
the former and hence contribute to the quadratic divergence. A minimally extended scalar sector 
is enough if one is interested only in the fine-tuning of the SM Higgs; however, one would naturally 
expect to satisfy the VC for the new scalars too. If the new scalars do not couple to the SM fermions
(like the singlet extension), one has to bring in some new fermions at the same time. 

In this paper, we concentrate on the extension of the SM with a complex triplet scalar 
\cite{gunion,paramita}. Why a scalar extension? As we will show, cancellation of quadratic
divergences to the Higgs mass requires extra bosonic degrees of freedom that couple to the 
SM Higgs at the tree-level. Extra gauge fields can also be invoked, but one anyway needs 
more scalars to give them mass in a gauge-invariant way. An alternative option, the two-Higgs 
doublet models, has been discussed elsewhere \cite{indrani-doublet}. 

Triplet scalars have received a lot of attention in the literature, including a detailed study of
couplings and mass spectrum \cite{logan}, radiative corrections, renormalizability issues, and 
precision observables \cite{radcor}, enhancement of the $h\to\gamma\gamma$ branching ratio 
\cite{arhrib}, and collider studies \cite{englert} \footnote{Triplet scalars may also be embedded 
in a bigger theory, like supersymmetry, vector fermions, or more scalar multiplets.}. 
However, their main appeal lies in neutrino mass
generation through the see-saw mechanism \cite{type2} 
with a lepton number ($L$) violating interaction, and also the type-II leptogenesis 
scenario \cite{leptogenesis}. As the 
complex triplet can couple to left-handed leptons to generate Majorana masses for the neutrinos
through $\Delta L = 2$ terms, there is no need to introduce any additional fermions in the model.  
The stability and unitarity conditions of such triplet models in the light of a 125 GeV Higgs
boson have been discussed in Ref.\ \cite{dilip}.

The vacuum expectation value (VEV) of the triplet is, of course, restricted from the 
$\rho$-parameter to be at most of a few GeV \cite{chen}. However, 
%
it is more than enough to generate the neutrino masses if the corresponding Yukawa 
couplings are of order unity. This is why we do not consider the extension of the SM 
with one complex and one real triplet, keeping the custodial SU(2) intact, which may 
give a large VEV for the neutral triplets \cite{gunion,georgi}. The mixing between the 
triplet and doublet states is proportional to the triplet VEV, which, being tiny, makes the 
mixing small too \cite{doub_trip_mix1}. Thus, the 125 GeV scalar is almost a pure doublet, which is 
completely consistent with its production cross-section and decay branching ratios. 

We will show that the introduction of a complex triplet can successfully address the naturalness 
problem for the doublet. Further multiplets, triplets or otherwise, might also help, but here we will try to 
keep the life simple by considering only the minimal extension, and that too without introducing any extra 
fermions. One notes that as the number of scalars 
increases, there is a compulsion to apply the naturalness condition to all of them, unless some of them 
happen to be extremely heavy (in which case they get frozen and do not contribute to the radiative 
corrections at a low energy). For the triplet scalars, the
naturalness problem is addressed through its coupling to 
the leptons \footnote{So, if necessary, one can keep the triplets light, but heavy triplets can easily be 
accommodated.}. One has also to take into account the stability conditions of the scalar potential. As will 
be seen, the potential of this model becomes unstable at a high energy scale; the scale depends on the 
initial choice of parameters but is at a few thousand TeV. One might argue that the potential could be made 
stable with higher-order corrections; even then, some of the couplings grow large and hit a Landau pole 
somewhere below $10^5$ TeV, which indicates the maximum energy where some NP must supersede the effective 
theory \footnote{Thus, the fine-tuning of Higgs mass is never more severe than 1 in 1000, which might not seem 
too bad, but we want to have a cancellation even less severe.}. 
We find out the parameter space for such a triplet-enhanced SM consistent with the VC 
as well as the stability of the scalar potential.

The paper is arranged as follows. In Section II, we show the complete scalar potential, the corresponding 
Veltman conditions, and the one-loop renormalization group (RG) equations for all the relevant couplings. 
In Section III, we study the RG evolution of the couplings
and its possible consequences. Section IV is on the scalar spectrum of such a model. 
We summarize and conclude in Section V.

\section{The Veltman Condition}

In the SM, with the scalar potential of the form 
\be
V(\Phi) = -\mu^2\Phi^\dag\Phi + \lambda\left( \Phi^\dag\Phi\right)^2\,,
\ee
the Higgs boson self-energy receives a quadratically divergent correction
\be
\delta m_h^2 = \frac{\Lambda^2}{16\pi^2} \left(6\lambda + \frac34 g_1^2 + \frac94 g_2^2
- 6 g_t^2\right)\,,
\label{smvc}
\ee
where $g_1$ and $g_2$ are the U(1)$_Y$ (not GUT-normalized) and SU(2)$_L$ gauge couplings 
respectively, and $g_t = \sqrt{2} m_t/v$ is the top quark Yukawa coupling. We treat all 
other fermions as massless, and use the cut-off regularization, $\Lambda$ being the cutoff scale.
The Veltman condition (VC) demands that the quantity inside the parentheses in Eq.\ (\ref{smvc})
be made zero, or at least controllably small,
 by some symmetry. There are further quadratic divergences coming from two-loop 
diagrams, but they are suppressed from one-loop contributions by a factor of 
$\ln(\Lambda/\mu)/16\pi^2$, where $\mu$ is the regularization scale, and we will neglect them 
here. 

One can say that the quadratic divergence is under control if, say, $|\delta m_h^2| \leq m_h^2$, 
which translates into \footnote{The fine-tuning condition is, of course, subjective, 
and one can easily allow a higher 
fine-tuning, but any fine-tuning defeats the motivation of the Veltman condition.}
\be
\left\vert m_h^2 + 2m_W^2 + m_Z^2 - 4m_t^2\right\vert \leq \frac{16\pi^2}{3} \frac{v^2}{\Lambda^2} m_h^2\,.
\label{smvc2}
\ee
This inequality is satisfied in the SM only for $v^2/\Lambda^2 \geq 0.1$, or $\Lambda \leq 760$ GeV,
which means that we should expect a NP at this scale. This, however, is almost ruled out by the LHC.
Eq.\ (\ref{smvc}) also shows that one needs a bosonic contribution to satisfy the Veltman condition. 

Let us now enhance the scalar sector with a complex triplet $X$, with weak hypercharge $Y=2$. 
The VEVs are  
\be
\bra\phi^{0}\ket = \frac{v_1}{\sqrt{2}}\,,\ \ 
\bra X^{0}\ket = v_2\,.
\ee
We can express the triplet in a bidoublet notation:
\be
X=\begin{pmatrix} X^+/\sqrt{2} & X^{++}\cr
X^0 & -X^+/\sqrt{2}\end{pmatrix}\,,
\ee
and the generic form of the $\Delta L = 2$ terms is 
\be
V_{\Delta L = 2} = - i f_{ab} L_a^T C^{-1} \tau_2 X L_b + {\rm h.c.}\,,
\ee
where $C$ is the charge conjugation operator, and $L = (\nu \ \ \ell)^T$ is 
the left-handed lepton doublet. If there is no leptonic flavor-changing 
neutral current, we can take the Yukawa coupling $f_{ab}$ to be diagonal. For subsequent 
discussion, we will not only assume $f_{ab}$ to be diagonal but also to be a 
multiple of the unit matrix: $f_{ab} = f \delta_{ab}$. While this seems to be at variance 
with the neutrino data, any form that correctly reproduces the neutrino masses and mixing
hardly changes our conclusions \footnote{For normal hierarchy, only one of the Yukawa 
couplings is large and the other two can be neglected; for inverted hierarchy, we have to keep
two equally large couplings and neglect the third one. Off-diagonal elements are to be introduced 
in $f_{ab}$ to generate the mixing angles. Anyway, a detailed discussion of neutrino mass 
matrix is outside the scope of this paper.}.

The scalar potential can be written as \cite{paramita}
\be
V = V_2 + V_3 + V_4\,,
\ee
where the individual terms are
\bea
V_2 &=& -\mu_1^2(\Phi^\dag\Phi) + \mu_2^2(X^\dag X)\,,\nonumber\\
V_3 &=& -a_0 (\Phi \Phi X^\dag) + h.c.\,,\nonumber\\
V_4 &=& \lambda_1 (\Phi^\dag\Phi)^2 + \lambda _2(X^\dag X)^2 + \lambda_3 (\Phi^\dag\Phi)(X^\dag X)
      + \lambda_4 (\Phi^\dag \tau_i \Phi)(X^\dag t_i X) + \lambda _5 \left\vert X^T \tilde{C} X\right\vert ^2\,,
       \label{onesing3}
\eea
with
\be
\tilde{C} = \begin{pmatrix} 0 & 0 & 1 \cr
0 & 1 & 0 \cr
1 & 0 & 0 \end{pmatrix}
\ee
and $\tau_i$s and $t_i$s ($i=1$--$3$) are the $2\times 2$ and $3\times 3$ Pauli matrices respectively, 
with $t_1 = \delta_{i,i+1} + \delta_{i,i-1}$, $t_2 = -i(\delta_{i,i+1} - \delta_{i,i-1})$, and $t_3 = 
{\rm diag}(1,0,-1)$.
Note that the triplet has a ``right-sign'' mass term, which ensures that the triplet VEV will arise only through 
the trilinear and quartic interactions, and can remain small without necessarily keeping the triplet light and 
hence jeopardizing the experimental constraints \footnote{The trilinear term can be banished by invoking 
discrete symmetries $\Phi\to-\Phi$ and $X\to -X$, but the latter also forbids the $\Delta L = 2$ terms.}. 
Without the trilinear term, there is a global O(2)
symmetry in the neutral scalar sector, so that there will be a physical Goldstone boson in the spectrum if
both neutral fields acquire VEV. One needs $a_0 > 0$ to prevent tachyonic mass of the 
scalars. Further ramifications of the trilinear term can be found in \cite{paramita}.

In terms of the real components, the fields can be written as 
\bea
\phi^0=\frac{1}{\sqrt{2}} (\phi^{0R} + v_1 + i\phi^{0I})\,,&&
\phi^{+} = \frac{1}{\sqrt{2}}(\phi_1+i \phi_2)\,,\nonumber\\
X^{0}=\frac{1}{\sqrt{2}} (X^{0R} + \sqrt{2}v_2 + i X^{0I})\,, &&
X^{++}=\frac{1}{\sqrt{2}}(X_1+i X_2)\,, \ \ 
X^{+}=\frac{1}{\sqrt{2}}(X'_1+i X'_2)\,,
\eea
where the neutral components have been vacuum-shifted. Only the terms in $V_4$ are relevant for computing 
quadratic divergences, so we rewrite those terms as \footnote{We correct a few sign mistakes in 
\cite{paramita}.}

\bea
V_4 &=& \frac14\lambda_1 \left[\left(\phi_1^2+\phi_2^2 + {\phi^{0R}}^2+{\phi^{0I}}^2\right)^2 \right]
+ \frac14\lambda_2\left[ \left( X_1^2+X_2^2 + {X_1^{'}}^2+{X_2^{'}}^2 + {X^{0R}}^2+{X^{0I}}^2\right)^2\right]\nonumber\\
&+&\frac14 \lambda_3 \left[ \left(\phi_1^2+\phi_2^2 + {\phi^{0R}}^2+{\phi^{0I}}^2\right) \left(X_1^2+X_2^2+
{X_1^{'}}^2+{X_2^{'}}^2 + {X^{0R}}^2+{X^{0I}}^2\right)\right]\nonumber\\
&+& \frac14\lambda_4\left[\left(\phi_1^2+\phi_2^2\right) \left(X_1^2+X_2^2\right) -
\left({\phi^{0R}}^2+{\phi^{0I}}^2\right)\left(X_1^2+X_2^2\right)
-\left(\phi_1^2+\phi_2^2\right)\left({X^{0R}}^2+{X^{0I}}^2\right)\right.\nonumber\\
&& + \left({\phi^{0R}}^2+{\phi^{0I}}^2\right)\left({X^{0R}}^2+{X^{0I}}^2\right) 
+ \sqrt{2} \left\{ \left(\phi_1+i\phi_2\right)\left(X_1^{'}+iX_2^{'}\right)
\left(X_1-iX_2\right)\left({\phi^{0R}}-i{\phi^{0I}}\right) + {\rm h.c.} \right\} \nonumber\\
&&\left.  + \sqrt{2} \left(\phi_1+i\phi_2\right)\left(X_1^{'}-iX_2^{'}\right)
\left({\phi^{0R}}-i{\phi^{0I}}\right)\left({X^{0R}}-i{X^{0I}}\right) + {\rm h.c.}\right]\nonumber\\
&+& \lambda_5 \left[ \left(X_1^2+X_2^2\right)\left({X^{0R}}^2+{X^{0I}}^2\right)+\frac{1}{4}
\left({X_1^{'}}^2+{X_2^{'}}^2\right)^2 \right.\nonumber\\
&& \left.
 + \frac{1}{2}\left(X_1^{'}+iX_2^{'}\right)\left(X_1^{'}+iX_2^{'}\right)
\left(X_1-iX_2\right)\left({X^{0R}}+i{X^{0I}}\right)+ {\rm h.c.}\right]\,.
\eea

With the triplet, the VC for the SM Higgs is modified to
\be
\delta m_h^2 = \frac{\Lambda^2}{16\pi^2} \left(6\lambda_1 + 3\lambda_3+ \frac34 g_1^2 + \frac94 g_2^2
- 6 g_t^2 \right)\,,
\label{tripletvc1}
\ee
With $m_h = 125$ GeV, $m_W = 80.41$ GeV, $m_Z = 91.19$ GeV, and $m_t = 174$ GeV, this fixes 
$\lambda_3 \approx 1.39$. This is large but still within the perturbative limit of $4\pi$. With 
$N$ identical triplets, $\lambda_3 \approx 1.39/N$.

The stability conditions of the scalar potential read
\be
\lambda_1, \lambda_2 \ge 0\,,
\lambda_2+2 \lambda_5\ge 0\,,
\lambda_3 \pm \lambda_4\ge -2\sqrt{\lambda_1 \lambda_2}\,, 
\label{stab7}
\ee
plus some other conditions that are not independent of these. Note that $\lambda_4$ and $\lambda_5$ can be negative. 
As we will show later, the lighter CP-even neutral state at 125.8 GeV is almost a pure doublet, which fixes $\lambda_1
\sim 0.13$. Thus, the stability conditions give a range for allowed values of $\lambda_4$ and a lower limit on $\lambda_5$ 
for any given value of $\lambda_2$. 
The VC for the triplet, which couples to the leptons through $\Delta L = 2$ interaction, reads
\be
\delta m_X^2 = \frac{\Lambda^2}{16\pi^2}\left(4\lambda_2+\lambda_3+2 \lambda_5+\frac12g_1^2 + g_2^2-3f^2\right)\,.
\label{tripletvc3}
\ee
Without the Yukawa term, $\delta m_X^2$ can never be made to vanish, even with possible negative values of $\lambda_5$, 
due to the stability conditions. There is no contribution proportional to $\lambda_4$ in Eq.\ (\ref{tripletvc3}); 
the quadratically divergent contributions cancel out. Also, even in the limit $\lambda_2, \lambda_5 \to 0$, the 
large value of $\lambda_3$ necessitates correspondingly large value of the Yukawa coupling $f$ 
($\sim {\cal O}(1)$) and hence an extremely 
tiny triplet VEV $v_2$ ($\sim {\cal O}(10^{-3}$ eV)), 
completely consistent with the $\rho$-parameter bound, as well as to the identification of the 
125 GeV resonance as the almost-pure SM doublet. The $3f^2$ term in Eq.\ (\ref{tripletvc3}) appears because of universal 
leptonic Yukawa couplings. For normal (inverted) hierarchy, we expect $3f^2\approx f_{\rm normal}^2 (2f_{\rm inverted}^2)$.

We would, of course, like the VCs for both the doublet and the triplet to be stable over the range of validity 
of the theory. We do not expect the VC combinations to remain exactly zero, because higher-order effects were not 
taken into account, but we would like a more or less stable behaviour 
\footnote{In a generic Yukawa theory, if the Higgs mass correction at one-loop remains zero at all scales, the leading 
two-loop quadratic corrections also vanish \cite{einhorn}.}. The one-loop renormalization group (RG) equations 
for the couplings are as follows:

\bea
16\pi^2 \beta_{\lambda_1} &=& 12\lambda_1^2 + \frac32\lambda_3^2 + \lambda_4^2 + 6g_t^2\lambda_1 - \frac32\lambda_1
\left(g_1^2+3g_2^2\right) -3g_t^4
 + \frac{3}{16}(g_1^4 + 2g_1^2g_2^2 +3g_2^4)\,,\nonumber\\
16\pi^2\beta_{\lambda_2} &=& 14\lambda_2^2 + \lambda_3^2 + \lambda_4^2 + 8\lambda_5^2 + 8\lambda_2\lambda_5 +  
2f^2\lambda_2 - 6\lambda_2 \left(g_1^2 + 2 g_2^2\right) + \frac32\left(2g_1^4 + 3 g_2^4 + 4g_1^2g_2^2\right) 
- f^4\,,\nonumber\\
16\pi^2\beta_{\lambda_3} &=& 6\lambda_1\lambda_3 + 8\lambda_2\lambda_3
 + 4\lambda_3\lambda_5 + 2\lambda_3^2 + 2\lambda_3(f^2 + 3g_t^2)+ \frac32 g_1^4 + 3 g_2^4 - \frac{15}{2}
\lambda_3 g_1^2 -\frac{33}{2}\lambda_3 g_2^2\,,\nonumber\\
16\pi^2\beta_{\lambda_4} &=& 2\lambda_1\lambda_4 + 2\lambda_2\lambda_4 - 4\lambda_4\lambda_5 + 4\lambda_3\lambda_4 + 
2\lambda_4^2 + 2\lambda_4(f^2+3g_t^2) - \frac{15}{2}g_1^2\lambda_4 - \frac{33}{2}g_2^2\lambda_4 + 3g_1^2g_2^2\,,\nonumber\\
16\pi^2\beta_{\lambda_5} &=& 12\lambda_2\lambda_5 + 2\lambda_5^2 - \lambda_4^2 + 2f^2\lambda_5 -6\lambda_5g_1^2 -
12\lambda_5g_2^2 + \frac32g_2^4 - 6g_1^2g_2^2 + \frac12 f^4\,,\nonumber\\
16\pi^2\beta_{f} &=& 6f^3 -\frac14 f \left(3g_1^2+9g_2^2\right)\,,
\label{all-rge}
\eea
where $\beta_h\equiv dh/dt$, and $t \equiv \ln(q^2/\mu^2)$, $\mu$ being the regularization scale.  
Note that our definition of $t$ differs by a factor of 2 from that used by some authors.

\section{Analysis}

To ensure that the VC for the doublet scalar is respected, one needs to fix only the value of $\lambda_3 
\approx 1.39$. 
Rest of the couplings 
are free parameters of the theory, except that Eq.\ (\ref{tripletvc3}) provides a relationship between 
$\lambda_2$, $\lambda_5$, and $f$. The only constraint on $\lambda_4$ comes from the stability condition. 
We, of course, assume all couplings to be perturbative ($\leq 4\pi$) 
over the entire range of validity of the theory. 

\begin{figure}[!htbp]
\begin{center}
{
\epsfxsize=7.5cm\epsfbox{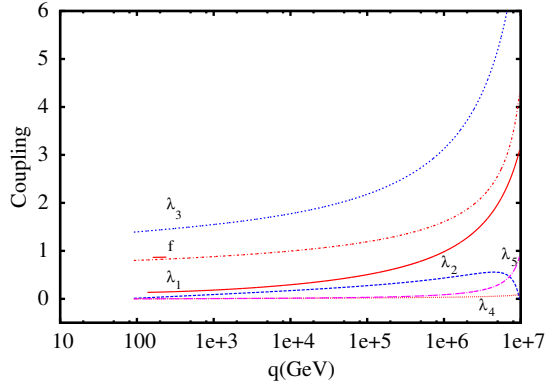}
}
\caption{The running of the couplings, using one-loop RG equations. The values at $q^2 = m_Z^2$ are 
$\lambda_2 = 0.01$, $\lambda_4 = \lambda_5 = 0$, and the rest are fixed by physical masses and/or Veltman 
Conditions.}
\label{fig:couprg}
\end{center}
\end{figure}

\begin{figure}[!htbp]
\begin{center}
{
\epsfxsize=7.5cm\epsfbox{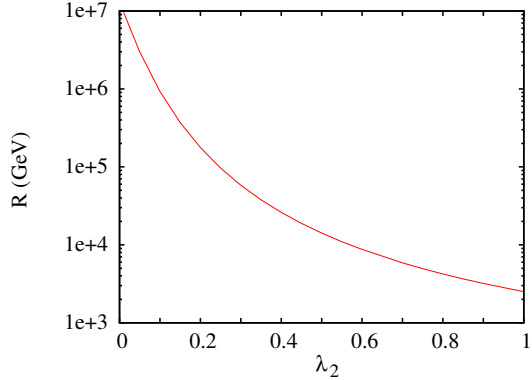}
}
\caption{The range of validity of the theory as a function of initial values of 
$\lambda_2$ keeping $\lambda_4 = 
\lambda_5 =0$.}
\label{fig:validity1}
\end{center}
\end{figure}

A scan over the free couplings is needed because their initial values, 
consistent with the stability conditions, fix the range of validity ${\cal R}$ of the theory. 
This is particularly true for $\lambda_2$. Over the entire parameter space $\lambda_2$ initially 
increases, and then reverses and becomes negative, indicating some other new physics
\footnote{One must remember that we are using only one-loop RG equations. However, the drop of $\lambda_2$ 
is so sharp, thanks to the rapidly increasing value of $f$, that we do not expect a qualitative change 
in the outcome even if we include higher-order terms.}. A typical evolution is shown in Fig.\ \ref{fig:couprg}. 
The reason for such a turning behaviour of $\lambda_2$ 
is easy to follow from the RG equations. The value of $f$ at the electroweak 
scale is fixed by the triplet VC, 
\be
f = \sqrt{(8\lambda_2+ 2\lambda_3+4 \lambda_5+ g_1^2+ 2g_2^2)/6}\,,
\ee
and this keeps the $\beta_{\lambda_2}$ positive. However, with increasing $q^2$, the Yukawa coupling $f$ 
increases so rapidly that the $-f^4$ term causes $\lambda_2$ to turn back, and ultimately the theory 
becomes unstable. The range ${\cal R}$ as a function of $\lambda_2$, keeping $\lambda_4 = \lambda_5  = 0$, 
is shown in Fig.\ \ref{fig:validity1}.

\begin{figure}[!htbp]
\begin{center}
{
\epsfxsize=7.5cm\epsfbox{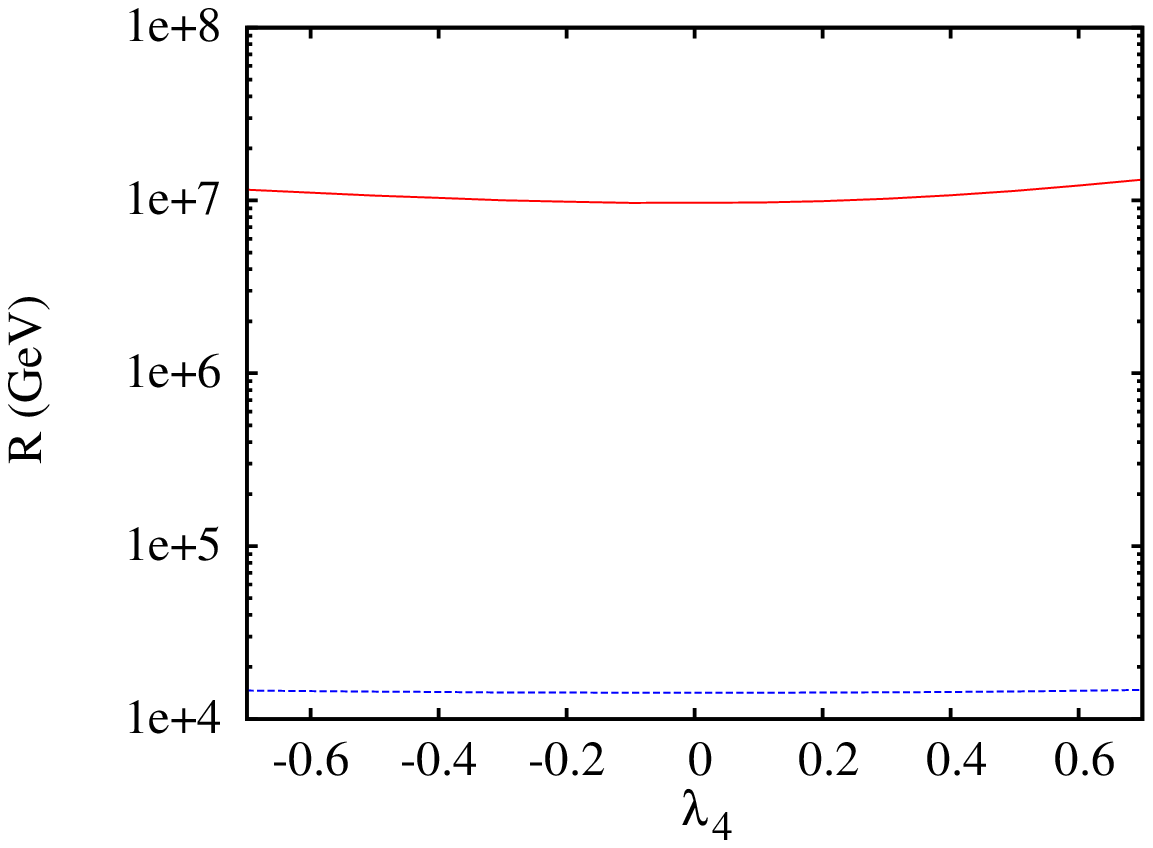}
}
{
\epsfxsize=7.5cm\epsfbox{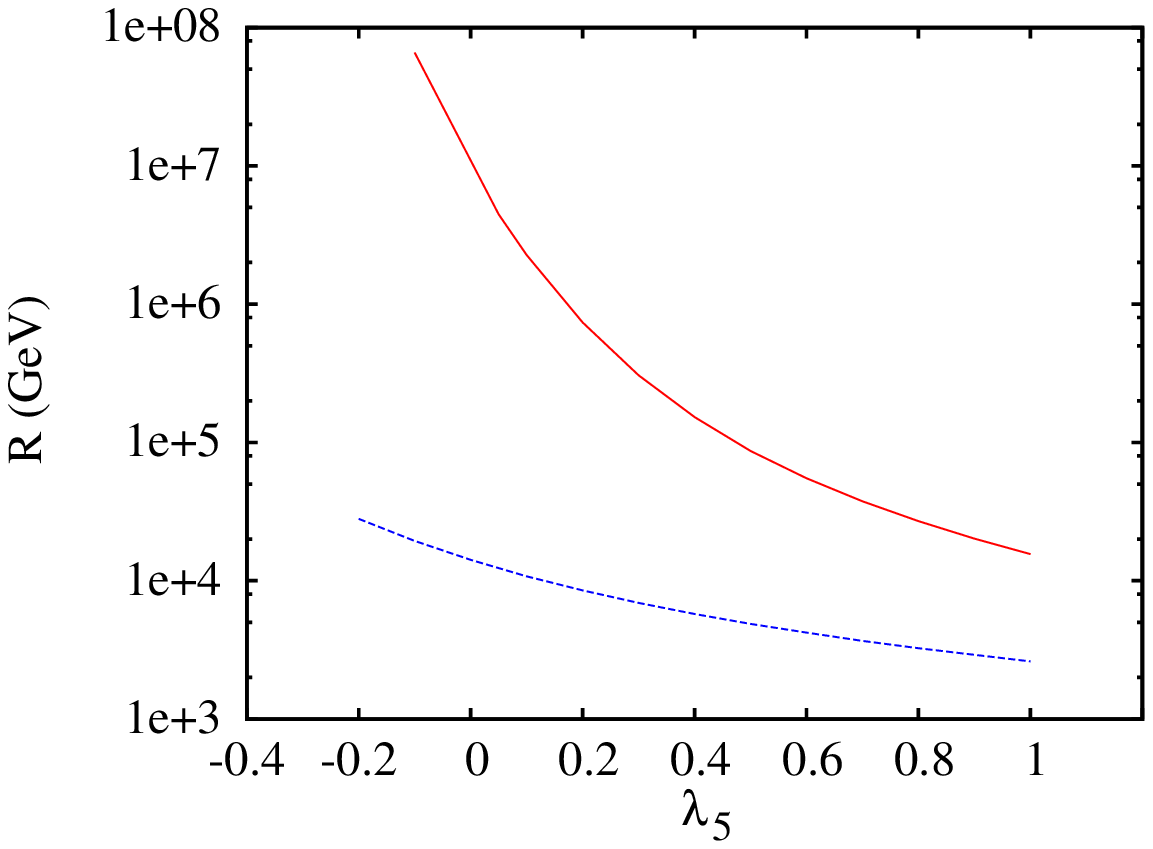}
}
\caption{Left panel: Range of validity as a function of initial values of $\lambda_4$, with 
$\lambda_5 = 0$, and $\lambda_2 = 0.01 (0.5)$ for upper red (lower blue) line. 
Right panel: The same as a funcion of initial values of $\lambda_5$, with $\lambda_4 = 0$, and 
$\lambda_2 = 0.01 (0.5)$ for upper blue (lower red) line.}
\label{fig:validity2}
\end{center}
\end{figure}

Fig.\ \ref{fig:validity1} might seem counter-intuitive; with increasing $\lambda_2$, $\beta_{\lambda_2}$
starts out from a more positive value, but ${\cal R}$ appears to shrink. This is because larger values 
of $\lambda_2$ need correspondingly larger values of $f$ to satisfy the triplet VC, and thus the turning 
of $\lambda_2$ occurs at a lower energy scale. Thus, we do not envisage $\lambda_2$ to be very large.

${\cal R}$ also depends on the initial values of $\lambda_4$ and $\lambda_5$, as shown in Fig.\ 
\ref{fig:validity2}. With increasing $|\lambda_4|$, the range increases. This is easy to understand; 
$\beta_{\lambda_2}$ picks up another positive contribution, $\lambda_4^2$, which keeps $\lambda_2$ 
positive for higher values of $q^2$. For $\lambda_5$, the deciding factor is the initial value of $f$; 
the lower the starting value of $f$, the higher the range of validity. 

All the other quartic couplings except $\lambda_2$ hit the Landau pole almost simultaneously, 
because of the coupled nature of the RG equations. This, however, occurs beyond ${\cal R}$ but 
typically between (2-4)${\cal R}$. Thus, the fine-tuning problem is never as severe as that of the SM.

\begin{figure}[!htbp]
\begin{center}
{
\epsfxsize=7.5cm\epsfbox{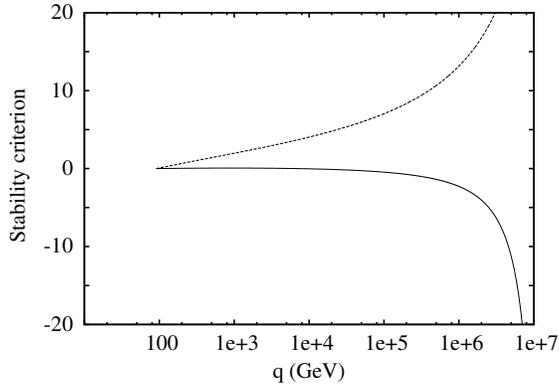}
}
\caption{Stability of the VC: the lower red line is for the triplet VC and the upper blue line is for the 
doublet VC. On the $y$-axis we plot $\delta m_{h(X)}^2 (16\pi^2/\Lambda^2)$. Drawn for $\lambda_2 = 0.01$,
$\lambda_4 = \lambda_5 = 0$.}
\label{fig:stability}
\end{center}
\end{figure}

As a last thing, we show, in Fig.\ \ref{fig:stability}, how the radiative corrections behave as we
go up the energy scale. What is plotted is $\delta m_{h,X}^2$ times $16\pi^2/\Lambda^2$, or in other 
words, the combinations of the couplings in Eqs. (\ref{tripletvc1}) and (\ref{tripletvc3}) 
as a function of the energy scale $q$. We find that 
the doublet VC is more or less stable while the triplet VC shows a sharp drop because of the steep 
increase in $f$. 

\section{Scalar Spectrum}

Let us first note that there exists a strong hierarchy between $v_1$ and $v_2$; $v_2/v_1 \sim 
{\cal O}(10^{-14})$. This has nothing to do with fine-tuning; it is but a reflection of the 
hierarchy between neutrino mass and the electroweak scale. 

The doubly-charged scalar $H^{++}$ is a pure triplet and its mass can directly be read off 
from Eq.\ (\ref{onesing3}):
\bea
m_{H^{++}}^2 &=& \mu_2^2 + \frac12 (\lambda_3-\lambda_4) v_1^2 + (2\lambda_2 + 4\lambda_5) 
v_2^2 \nonumber\\
& = & 4 \lambda_5 v_2^2 - \lambda_4 v_1^2 + \frac{v_1^2}{2}\frac{a_0}{v_2}\,.
\label{mass-doub}
\eea
There are two singly-charged fields. After diagonalizing the mass matrix, one of them turns 
out to be the Goldstone boson (which, in the limit $v_2 \ll v_1$, is almost a pure doublet),
and the other has a mass
\be
m_{H^+}^2 = \frac12 \left(v_1^2+4v_2^2\right) \left(\frac{a_0}{v_2}-\lambda_4\right)\,.
\label{mass-sing}
\ee
To get Eqs.\ (\ref{mass-doub}) and (\ref{mass-sing}), we have used the minimization 
conditions \cite{paramita}:
\bea
-\mu_1^2 + v_1^2\lambda_1 + v_2^2(\lambda_3+\lambda_4) &=& 2 a_0 v_2\,,\nonumber\\
\mu_2^2 + 2v_2^2\lambda_2 + \frac12 v_1^2(\lambda_3+\lambda_4) &=& \frac{a_0v_1^2}{2 v_2}\,.
\eea
If $a_0/v_2 \gg 1$, $H^+$ and $H^{++}$ are almost mass-degenerate, and their masses can be 
large; the quartic couplings hardly have any effect on their masses. 

The CP-odd neutral scalar, $A$, is again almost entirely the triplet component $X^{0I}$, whose
mass is given by
\be
m_A^2 = \frac12 \frac{a_0}{v_2} \left(v_1^2+8v_2^2\right)\,.
\ee
Thus, not only $A$ is almost degenerate with $H^+$ and $H^{++}$ in the limit $a_0/v_2\gg 1$,
$a_0$ has to be nonzero in order to prevent the Goldstone boson \cite{paramita} and hence 
$v_2$ must be nonzero, albeit small, for the theory to be consistent.

The mass matrix for CP-even neutral scalars can be written, with the help of the 
minimization conditions of the scalar potential, as
\be
{\cal M}^{0R}=\begin{pmatrix}
               v_1^2 \lambda_1 & \frac{1}{\sqrt{2}} v_1 v_2 \varphi \\
               \frac{1}{\sqrt{2}} v_1 v_2 \varphi & 2 v_2^2 \lambda_2 + \frac14 \frac{a_0v_1^2}{v_2}
              \end{pmatrix}
\ee
where $\varphi = \lambda_3+\lambda_4 - a_0/v_2$. This is almost a diagonal matrix for $v_2\ll v_1$,
so that $m_h^2 = 2\lambda_1 v_1^2$. 
Apart from the 125 GeV scalar, all the other scalars are (almost) pure triplet and close to
degenerate for $a_0/v_2 \ll 1$. The charged 
scalars can be pair produced at the LHC, through $\gamma$ or $Z$ exchange. Single production is
suppressed by the tiny value of $v_2$. Once 
produced, they will dominantly decay into a lepton pair, irrespective of their mass. 
This is in contrast to the case where $v_2$ is sizable and di-gauge decay channels may be 
dominant. Such dilepton signals
from $H^{++}$ have been looked for by both ATLAS and CMS collaborations \cite{atlas-dilep},
and a bound of $m_{H^{++}} \gtap 400$ GeV has been established. This translates into $a_0/v_2 
\gtap 5.3$. 

Thus, the main effect of the Veltman condition for the triplets is to enforce a Yukawa 
coupling $\sim {\cal O}(1)$ and hence a tiny value of $v_2$. This makes the triplet decouple
from the doublet, for all practical purpose, unless the dimensionless quantity $a_0/v_2$ 
falls significantly below the ATLAS and CMS limits. It also makes the triplet scalars almost 
mass degenerate. 
Consequently, the only significant production channel is through an $s$-channel $\gamma$ or $Z$ 
exchange. While $a_0/v_2 > 4
\lambda_1\approx 0.5$ ensures that the lighter CP-even neutral scalar is the doublet, even light triplets
are going to be missed unless they can be pair produced. 

\section{Summary}

The SM, as it stands, is definitely not enough to address the fine-tuning problem.
If we want to make a minimalistic extension of the SM to address the fine-tuning problem of the 
Higgs mass, the new degrees of freedom have to be bosonic. 

Extension of the SM by scalars demands that the fine-tuning problem of all the scalars be addressed 
simultaneously, unless some of them are extremely heavy. While some of the scalar couplings can in 
principle be negative, stability of the scalar potential forces the new scalars to have some fermionic 
couplings. In this respect, a complex triplet is an interesting alternative as (i) it can couple to 
the SM leptons through $\Delta L = 2$ interactions and generate Majorana masses for the neutrinos; 
(ii) the smallness of the neutrino masses ensures that the triplet VEV is tiny if the new Yukawa couplings 
are of order unity, so that the $\rho$-parameter constraint is easily evaded. Moreover, the lightest 
CP-even scalar remains an almost pure doublet, in conformity with the LHC Higgs data.

Addition of the triplet gives an extra positive contribution to the Veltman condition for the doublet.
The coupling $\lambda_3$, as defined in Eq.\ (\ref{onesing3}), turns out to be $1.39$ for exact 
cancellation of one-loop quadratic corrections (and $1.39/N$ if there are $N$ number of identical 
complex triplets). Similarly, with the help of other couplings, one can satisfy the triplet VC too. 

We have also checked the evolution of the couplings for the stability of the scalar potential, albeit 
at the one-loop level. The contribution of two-loop diagrams are suppressed by an additional factor of 
$\ln(\Lambda^2/m^2)/16\pi^2$, which is at most as a few per cent level to the one-loop contributions 
for $\Lambda \sim 10^6$ GeV. 
The potential becomes unstable as $\lambda_2$ becomes negative at some high scale
${\cal R}$ at the ballpark of thousands of TeV. This indicates some new physics at this scale which 
must change the $\beta$-functions. If we neglect this feature, the other scalar quartic couplings blow up
within one order of magnitude of ${\cal R}$, so some new physics is indicated anyway.

One might wonder about the motivation of introducing the Veltman condition to address the fine-tuning 
problem if the theory itself becomes invalid at, say, $10^6$ GeV. We would argue that it is still a 
useful approach; the fine-tuning is still there in the SM, maybe not as terrible as 1 in $10^{17}$ 
but even 1 in $10^4$ is bad enough, and should be addressed. 
At this point, we do not know what the nature of the NP at ${\cal R}$ is, but the theory below ${\cal R}$ 
can be treated as an effective theory, with those heavy degrees of freedom integrated out. In a subsequent 
publication, we will discuss the role of effective higher-dimensional operators to the Veltman condition.

\section{Acknowledgements}

We thank Nabarun Chakrabarty for pointing out an error in the original manuscript. 
I.C. acknowledges CSIR, Govt.\ of India, for a research fellowship. A.K. acknowledges DST, Govt.\ of India, 
and CSIR, Govt.\ of India, for research support. He also acknowledges the partial support from the DRS 
programme of UGC, Govt.\ of India.


\begin{thebibliography}{99}

\bibitem{veltman}
M.~J.~G.~Veltman,
  Acta Phys.\ Polon.\ B {\bf 12}, 437 (1981).


\bibitem{einhorn}
M.~B.~Einhorn and D.~R.~T.~Jones,
  Phys.\ Rev.\ D {\bf 46}, 5206 (1992).


\bibitem{aksrc}
 A.~Kundu and S.~Raychaudhuri,
  Phys.\ Rev.\ D {\bf 53}, 4042 (1996)
  [hep-ph/9410291].

\bibitem{drozd}
B.~Grzadkowski and J.~Wudka,
  Phys.\ Rev.\ Lett.\  {\bf 103}, 091802 (2009)
  [arXiv:0902.0628 [hep-ph]];\\
 A.~Drozd, B.~Grzadkowski and J.~Wudka,
  JHEP {\bf 1204}, 006 (2012)
  [arXiv:1112.2582 [hep-ph]];\\
 F.~Bazzocchi and M.~Fabbrichesi,
  Phys.\ Rev.\ D {\bf 87}, no. 3, 036001 (2013)
  [arXiv:1212.5065 [hep-ph]].

\bibitem{indrani}
I.~Chakraborty and A.~Kundu,
  Phys.\ Rev.\ D {\bf 87}, 055015 (2013)
  [arXiv:1212.0394 [hep-ph]].
  
\bibitem{masina}
I.~Masina and M.~Quiros,
  Phys.\ Rev.\ D {\bf 88}, 093003 (2013)
  [arXiv:1308.1242 [hep-ph]].

\bibitem{gunion}
J.~F.~Gunion, R.~Vega and J.~Wudka,
  Phys.\ Rev.\ D {\bf 42}, 1673 (1990);
  Phys.\ Rev.\ D {\bf 43}, 2322 (1991).

\bibitem{paramita}
P.~Dey, A.~Kundu and B.~Mukhopadhyaya,
  J.\ Phys.\ G {\bf 36}, 025002 (2009)
  [arXiv:0802.2510 [hep-ph]].
  
\bibitem{indrani-doublet} 
I.~Chakraborty and A.~Kundu,
  arXiv:1404.3038 [hep-ph].
  
\bibitem{logan}
I.~Gogoladze, N.~Okada and Q.~Shafi,
  Phys.\ Rev.\ D {\bf 78}, 085005 (2008)
  [arXiv:0802.3257 [hep-ph]];\\
H.~E.~Logan and M.~-A.~Roy,
  Phys.\ Rev.\ D {\bf 82}, 115011 (2010)
  [arXiv:1008.4869 [hep-ph]];\\
F.~Arbabifar, S.~Bahrami and M.~Frank,
  Phys.\ Rev.\ D {\bf 87}, 015020 (2013)
  [arXiv:1211.6797 [hep-ph]].

\bibitem{radcor}
M.~-C.~Chen, S.~Dawson and C.~B.~Jackson,
  Phys.\ Rev.\ D {\bf 78}, 093001 (2008)
  [arXiv:0809.4185 [hep-ph]];\\
M.~Aoki, S.~Kanemura, M.~Kikuchi and K.~Yagyu,
  Phys.\ Lett.\ B {\bf 714}, 279 (2012)
  [arXiv:1204.1951 [hep-ph]];
  Phys.\ Rev.\ D {\bf 87}, 015012 (2013)
  [arXiv:1211.6029 [hep-ph]];\\
T.~Fukuyama, H.~Sugiyama and K.~Tsumura,
  JHEP {\bf 1003}, 044 (2010)
  [arXiv:0909.4943 [hep-ph]];\\
S.~Kanemura and K.~Yagyu,
  Phys.\ Rev.\ D {\bf 85}, 115009 (2012)
  [arXiv:1201.6287 [hep-ph]].

\bibitem{arhrib}
A.~Arhrib, R.~Benbrik, M.~Chabab, G.~Moultaka and L.~Rahili,
  arXiv:1202.6621 [hep-ph];\\
A.~G.~Akeroyd and S.~Moretti,
  Phys.\ Rev.\ D {\bf 86}, 035015 (2012)
  [arXiv:1206.0535 [hep-ph]].

\bibitem{englert}
M.~Chaichian, P.~Hoyer, K.~Huitu, V.~A.~Khoze and A.~D.~Pilkington,
  JHEP {\bf 0905}, 011 (2009)
  [arXiv:0901.3746 [hep-ph]];\\
A.~G.~Akeroyd and C.~-W.~Chiang,
  Phys.\ Rev.\ D {\bf 80}, 113010 (2009)
  [arXiv:0909.4419 [hep-ph]];\\
S.~Godfrey and K.~Moats,
  Phys.\ Rev.\ D {\bf 81}, 075026 (2010)
  [arXiv:1003.3033 [hep-ph]];\\
M.~Aoki, S.~Kanemura and K.~Yagyu,
  Phys.\ Rev.\ D {\bf 85}, 055007 (2012)
  [arXiv:1110.4625 [hep-ph]];\\
C.~Englert, E.~Re and M.~Spannowsky,
  Phys.\ Rev.\ D {\bf 87}, no. 9, 095014 (2013)
  [arXiv:1302.6505 [hep-ph]];
  Phys.\ Rev.\ D {\bf 88}, 035024 (2013)
  [arXiv:1306.6228 [hep-ph]].

  
\bibitem{type2}
J.~Schechter and J.~W.~F.~Valle,
  Phys.\ Rev.\ D {\bf 22}, 2227 (1980);\\
G.~Lazarides, Q.~Shafi and C.~Wetterich,
  Nucl.\ Phys.\ B {\bf 181}, 287 (1981);\\
R.~N.~Mohapatra and G.~Senjanovic,
  Phys.\ Rev.\ D {\bf 23}, 165 (1981);\\
W.~Chao and H.~Zhang,
  Phys.\ Rev.\ D {\bf 75}, 033003 (2007)
  [hep-ph/0611323];\\
C.~-S.~Chen and C.~-M.~Lin,
  Phys.\ Lett.\ B {\bf 695}, 9 (2011)
  [arXiv:1009.5727 [hep-ph]];\\
A.~Chaudhuri, W.~Grimus and B.~Mukhopadhyaya,
  JHEP {\bf 1402}, 060 (2014)
  [arXiv:1305.5761 [hep-ph]].

  
\bibitem{leptogenesis}
See, {\em e.g.} 
E.~Ma and U.~Sarkar,
  Phys.\ Rev.\ Lett.\  {\bf 80}, 5716 (1998)
  [hep-ph/9802445];\\
T.~Hambye, E.~Ma and U.~Sarkar,
  Nucl.\ Phys.\ B {\bf 602}, 23 (2001)
  [hep-ph/0011192];\\
D.~Aristizabal Sierra, M.~Dhen and T.~Hambye,
  arXiv:1401.4347 [hep-ph].
  
\bibitem{dilip}
P.~S.~Bhupal Dev, D.~K.~Ghosh, N.~Okada and I.~Saha,
  JHEP {\bf 1303}, 150 (2013)
  [Erratum-ibid.\  {\bf 1305}, 049 (2013)]
  [arXiv:1301.3453].

\bibitem{chen}
T.~Blank and W.~Hollik,
  Nucl.\ Phys.\ B {\bf 514}, 113 (1998)
  [hep-ph/9703392];\\
M.~-C.~Chen, S.~Dawson and T.~Krupovnickas,
  Int.\ J.\ Mod.\ Phys.\ A {\bf 21}, 4045 (2006)
  [hep-ph/0504286];
  Phys.\ Rev.\ D {\bf 74}, 035001 (2006)
  [hep-ph/0604102];\\
S.~Banerjee, M.~Frank, and S.~K.~Rai, 
arXiv:1312.4249.
  
\bibitem{georgi}
H.~Georgi and M.~Machacek,
  Nucl.\ Phys.\ B {\bf 262}, 463 (1985);\\
M.~S.~Chanowitz and M.~Golden,
  Phys.\ Lett.\ B {\bf 165}, 105 (1985).


\bibitem{doub_trip_mix1}
J.~R.~Forshaw, D.~A.~Ross and B.~E.~White,
  JHEP {\bf 0110}, 007 (2001)
  [hep-ph/0107232].
  
\bibitem{atlas-dilep}
S.~Chatrchyan {\it et al.}  [CMS Collaboration],
  Eur.\ Phys.\ J.\ C {\bf 72}, 2189 (2012)
  [arXiv:1207.2666 [hep-ex]];\\
G.~Aad {\it et al.}  [ATLAS Collaboration],
  Eur.\ Phys.\ J.\ C {\bf 72}, 2244 (2012)
  [arXiv:1210.5070 [hep-ex]].


   

\end{thebibliography}
\end{document}